# Brownian Colloids in Optothermal Field: An Experimental Perspective

G.V. Pavan Kumar

Department of Physics, Indian Institute of Science Education and Research, Pune - 411008, INDIA

email: pavan@iiserpune.ac.in

web: http://sites.iiserpune.ac.in/~pavan/

## Table of Contents

# ABSTRACT

Colloidal matter undergoing Brownian motion serves as a model system to study various physical phenomena. Understanding the effect of external perturbation on the assembly and dynamics of 'Brownian colloids' has emerged as a relevant research issue in soft matter and biological physics. Optical perturbation in the form of photonic forces and torques has added impetus to this exploration. In recent years, optothermal effects arising due to optical excitation of mesoscale matter have expanded the toolbox of light-colloidal matter interactions. In this perspective, we present an experimental viewpoint on some of the developments related to the assembly and dynamics of Brownian colloids driven by the optothermal field. Furthermore, we discuss some interesting prospects on driven colloidal matter that can have implications on soft matter physics and soft photonics.

# INTRODUCTION

## Brownian Colloids, Optical and Optothermal Fields

A Brownian particle suspended in a fluid is a prototypical colloidal system to study random forces. The randomness in this process arises due to stochastic interactions between the particle and the molecules surrounding it.[1,2] The temperature and viscosity of the fluid influence the stochasticity in the Brownian dynamics. An optical trap based on a laser beam can be a model system for deterministic spatiotemporal force.[3]. If we bring an optical trap into the vicinity of the Brownian particle, then we can study the interaction between deterministic and random forces.[4]

Given that an optical trap can facilitate deterministic perturbations, we can tailor the interaction between the laser beam and the Brownian particles.[5–7] The nature of this interaction further depends on the characteristics of the laser beam and that of the particles. An unfocused laser imparts two kinds of forces onto a Brownian particle: optical gradient force and scattering force.[8,9] Varying certain parameters of the laser beam can control these two forces. Whereas optical gradient forces tend to bring the particle of interest into the center of the laser beam, the scattering forces push the particle away.[10] The interplay of push and pull forces makes the interaction sensitive to the properties of both the laser and the Brownian particle.[11,12]

Generally, a tightly focused laser beam is employed to create an optical trap.[13] This tight focusing condition ensures that the optical gradient force is more than the scattering force, thereby confining the Brownian particle to an 'optical potential.' Tight focusing of laser beams can also bring the vectorial nature of light into play and can facilitate a more complex landscape of forces.[14] Given that laser beams can carry linear and angular momentum states,[15,16] one can impart deterministic forces and torques on the Brownian particle. This facilitates translation and rotational degrees to be studied in this system. Such studies can offer models for a variety of dynamic processes, including quantum[17,18], biological[19] and ecological systems.[20,21]

Absorptive effects accompany the interaction of light with matter, but for most of the scattering interactions, the absorption can be negligible.[22] When the cross-section of absorption is significant, it cannot be neglected.[23] The absorption of light by colloidal systems facilitates non-radiative interaction channels.[22,24,25] These channels can further lead to optical heating. An increment in temperature created by these thermal fields can be controlled by optical excitation.[26] In recent times, the creation of the so-called optothermal fields[23,27] has been harnessed for colloidal assembly and transport.[28–34] This can be achieved either by directly illuminating light-absorbing colloids[35] or by utilizing the temperature gradients[36] created by optical heating of certain materials, such as metals. Specifically, the thermoplasmonic fields[37–39] facilitated by illuminating well defined metallic nanostructures have gained prominence due to developments in nanoscale chemical synthesis and top-down nano-fabrication methods. These developments in optothermal field generation have created utility in trapping, tweezing, transport and assembly of colloids and have expanded the toolbox of optical manipulation.[27,40–44] In this perspective article, we wish to communicate some developments emanating from our research.

## Structured Light as a Toolbox

Generalized solutions to the Helmholtz wave equation are rich in spatiotemporal patterns.[45–47] These patterns, either in two or three dimensions, can be explored in the optical domain to realize structured light in laser beams.[48,49] The structuring itself was not an easy task until new technology emerged in the past couple of decades.[50] Some of the emergent technology includes liquid-crystal-based spatial light modulators, Galvano mirrors, acoustic-optical modulators and deflectors, and related devices.[51,52] Due to the remarkable progress in pixel-level variation of spatial light modulators, the generation of structured light has become effective and straightforward. Thanks to this, we can electronically control SLMs via digital holograms and reconfigure the beam profile of lasers at high accuracy. This helps us to spatiotemporally modulate the intensity, phase, polarization and wavevector of laser beams in real time.[53]

As a result of this, reconfigurable optical potentials can be realized. One can transform a conventional Gaussian optical beam into a variety of configurations. Among the many varieties of optical beams, one can generate Laguerre-Gaussian, Hermite Gaussian, Bessel, Airy-Gaussian and many other configurations.[54–56] An interesting aspect is that one can now tailor the linear and angular momentum of laser beams using a variety of methods. Both the spin and orbital angular momentum states can be induced in a beam and further coupled with other parameters of light either in scalar or vectorial forms.[56] This 'structuring' of light has emerged as one of the fascinating research areas in optical physics that has transformed optical microscopy and nanoscopy, including optical and optothermal force transduction.

# BROWNIAN COLLOIDS – ASSEMBLY AND DYNAMICS

In this section, we will discuss some of our contributions and experimental perspectives related to the interaction of Brownian colloids with optical and optothermal fields.

## Evolutionary assembly of light-absorbing colloids

Temperature is an important environmental cue that can affect the assembly and dynamics of colloids in an optical trap. Understanding the spatio-temporal evolution of colloids is a pertinent way to study the effect of temperature distribution on the dynamics of soft matter. Within this framework, exploring how light-absorbing colloids behave when subjected to structured optical potentials is interesting. By comparing their behavior with conventional colloids, which do not absorb light significantly, the effect of distributed optothermal potential can be unveiled. This motivated us to experimentally study [35] spatio-temporal dynamics of light-absorbing colloids in a simple structured optical potential. We chose polystyrene colloids around 2 microns in diameter infused with iron oxide nanoparticles as light-absorbing colloids. To realize a symmetric optical potential, we created a defocused 532-nanometer laser beam through a high numerical aperture objective lens. Various configurations of colloids were studied, including single, double and multiple colloids confined to the optothermal potential (See Fig. 1a).

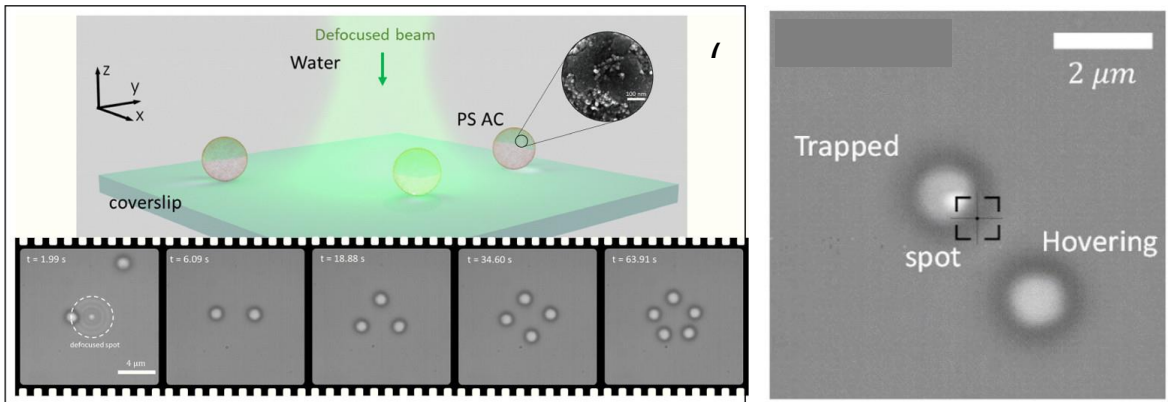

**Fig 1**: Evolutionary optical assembly of light-absorbing colloids (PS ACs – polystyrene colloids infused with iron oxide nanoparticles). (a) Illustration of the 2-dimensional trap using a defocused laser beam to create the assembly. Bottom row - Time series of the assembly of PS ACs under a defocused illumination. (b) Thermophoretic hovering of a PS AC. A heated trapped colloid at the focal spot results in temperature distribution in the surrounding medium. A second PS AC undergoes thermophoretic migration toward the heat center and eventually hovers near a trapped colloid at a certain distance. Reproduced with permission. [35]

Unlike conventional colloids, light-absorbing colloids get trapped off-center to the symmetric beam. This is because the trap center constitutes the region of maximum intensity, so the heating is maximum. This creates a short-range repulsion of the colloids close to the trap center. This acts against the long-range optical gradient force that still keeps the colloids confined in the trap. Next, we explored how light-absorbing colloidal pairs interact with a focused and defocused Gaussian optical trap. In the focused configuration, we observed the hovering behavior of colloids (see Fig.1b), which interacted with the colloid trapped close to the center. This unique behavior revealed how colloids could interact via a long-range optothermal interaction, which is generally absent in the case of conventional colloids. When a colloidal pair was realized in a defocused trap, we observed orientational sensitivity to the optical polarization of the trap. This indicated that polarization of the trap could be harnessed as a control parameter to assemble and modify the dynamics in an optothermal interaction.

 Further, we populated the defocused trap with multiple colloids and observed the evolutionary assembly of colloids in hexagonal symmetry. Unlike conventional colloids, light-absorbing colloids do not touch each other, thus exhibiting a spatially separated assembly unique to the optothermal interaction. This experiment indicates that optothermal colloidal interaction can lead to some unconventional assembly and can be used as test-beds to understand evolutionary assemblies in which short and long-range interactions work against each other.

# The active-passive colloidal mixture can lead to the emergence of directional rotation in an optical trap

Can we observe the emergence of new degrees of freedom in an optothermal trap where colloidal mixtures are confined to an optical trap? To address this prospect, we further explored[57] the question: what happens when we mix passive colloids with thermally active colloids in an optical trap? Specifically, we were interested in experimental probing of the emergent dynamics in such a trap.

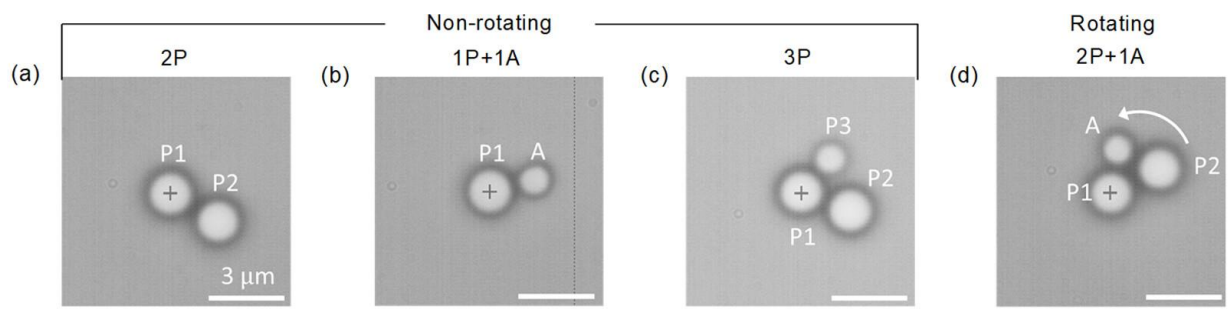

**Fig 2**: Dynamics of structures composed of passive and thermally active colloids. No rotational motion is observed in dimer structures composed of (a) two passive colloids (P1 and P2) and (b) one passive colloid (P1) and one thermally active PS colloid (A). (c) Passive trimer structure does not exhibit any rotational dynamics. (d) Active trimer composed of two passive colloids (P1, P2) and one thermally active colloid (A) exhibits counterclockwise rotation. Reproduced with permission.[57]

To implement this, we used a mixture of polystyrene colloids of 1.3 microns in diameter and mixed them with iron oxide-infused polystyrene colloids of a similar size. The infusion of iron oxide renders the colloid to be light-absorbing in nature and, therefore acts as a thermally active colloid. A two-dimensional optical trap created by a focused 532-nanometer Gaussian laser beam was realized by an optical tweezer at a high numerical aperture (0.95NA). A fast camera operating at 100 frames per second was employed to capture the dynamics.

There were some interesting observations from this experiment, and the following were the main features. We observed an emergence of directional rotation when there is an asymmetric mixture of active and passive colloids. The rotation critically depended on the geometrical asymmetry and the compositional heterogeneity (see Fig. 2d) and was absent for other configurations (see Fig.2a-c). The speed of rotation was controlled by modulating the power of the laser trap. Furthermore, the rotation speed also depended on the relative sizes of the active and passive colloids utilized in the experiment.

One of the main features of the experiment was that the handedness of the rotation could be controlled by placing the laser beam at a specific location with respect to the colloidal assembly. Interestingly, for the same configuration of the colloids, the rotation direction could be switched by changing the position of the laser trap relative to the colloidal assembly.

The origin of the rotation was mainly due to the torque induced by thermo-osmotic flow. Specifically, this thermo-osmotic flow depended on the asymmetric heating in the assembly created by the laser trap. The temperature gradient mainly depended on the geometrical arrangement of the constituent colloids. The interaction itself had multiple contributions, including a long-range optical force, which brought the colloids into the potential well, and a prominent thermophoretic force, which had direct implications on the internal dynamics within the trap. These forces further influence the hydrodynamic interaction of the colloids and their surroundings.

The main inference that we can draw from these observations is that the directional rotation can be induced and controlled by changing the parameters of the trap and the composition of the active and passive colloids. The mixture of active and passive colloids in an optical trap further facilitates unconventional dynamics, and such dynamics can lead to the emulation of some processes that are of relevance in natural systems. Given that the trap parameters can be used to control dynamics, one can engineer dynamical processes. This means one can employ the parameters of the laser trap to induce a rotational degree of freedom and can be utilized as test beds of Brownian rotors and rotational micro-engines.[58] There is further scope for exploration in this direction and we envisage interesting results to emerge.

# Plasmonic colloids in gold nanoparticle-driven optothermal traps

How does the optothermal field of a single gold nanoparticle influence the trapping potential in a fluid? This is a question that we tried to address in 2021[59], where we utilized a single gold nanoparticle-driven thermoplasmonic tweezer platform to accumulate and move gold nanoparticles in a fluid. Furthermore, having agglomerated a few gold nanoparticles, we were able to capture surfaced enhanced Raman scattering (SERS) signatures down to the single molecule limit.[59] This capability was facilitated thanks to the mechanism that was driven by surfactants in the solution, which Zhang and co-workers have extensively studied.[60]

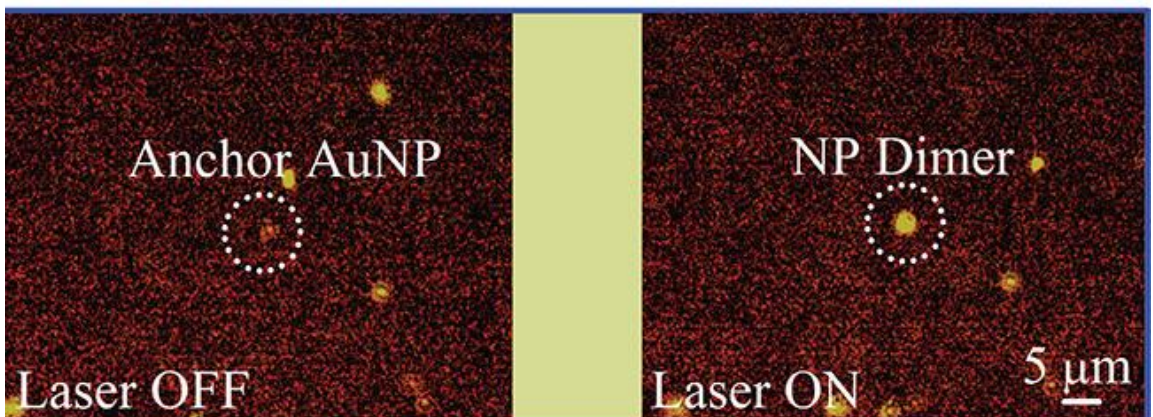

**Fig 3**: Trapping of a single gold nano-colloid in the vicinity of an anchored gold nanoparticle on a glass surface. A reversible dimer formation was observed by switching the laser ON and OFF. Anchored gold nanoparticle (AuNP): 150 nm, trapped plasmonic colloid: 250 nm. Reproduced with permission.[59]

One of the main advantages of this process is that you do not have to use intense laser excitations to trap single gold nanoparticles. This method overcomes the problem of laser damage and can be deployed in environments where the trapped object is sensitive to external perturbations. In our study[59], we performed systematic experiments and numerical analysis to show that even a single gold nanoparticle placed on a glass surface can facilitate optothermal potentials, which can be further utilized for trapping individual gold colloids (see Fig. 3). Furthermore, by using the 'bi-analyte' method[61,62], we were able to establish single-molecule SERS sensitivity and thus revealed its versatility both in terms of optothermal tweezing and Raman spectroscopy. An important offshoot of this technique is that you can also use this to temporarily form plasmonic dimers by creating optothermal potentials up to the limit of single nanoparticles. All these were achieved by utilizing laser powers as small as 0.1 milliwatt per micrometer square which is very nominal by conventional optical trapping standards of nanoparticles. Also, this technique can be used to trap a variety of plasmonic structures of various sizes, and our study revealed this capability for plasmonic silver and gold colloids in the range of 60 nanometers to 400 nanometers. These studies were followed by deterministic trapping and spectroscopy of single nanodiamonds[63] facilitated by the optothermal field facilitated by an anchored gold nanoparticle. We envisage that anchored-gold-nanoparticle-based trapping will find further utility in nano and bio-colloidal assembly and transport.

# Plasmonic Nanowires assist assembly and crystallization of colloids

Plasmonic nanoparticles can be effectively used to study thermoplasmonic effects. The localized plasmon resonance and absorption facilitate a localized heat source. As an extension, delocalized plasmons can be realized using certain geometries, such as silver nanowires. An interesting aspect of such wires is that the plasmons are delocalized and are sensitive to the excitation polarization. This means that the non-radiative component in the wire can also be influenced by optical polarization.[36] This further influences the temperature distribution in plasmonic structures, and it would be of relevance to study the assembly and dynamics of colloids under optothermal gradients achieved by plasmonic nanowires.

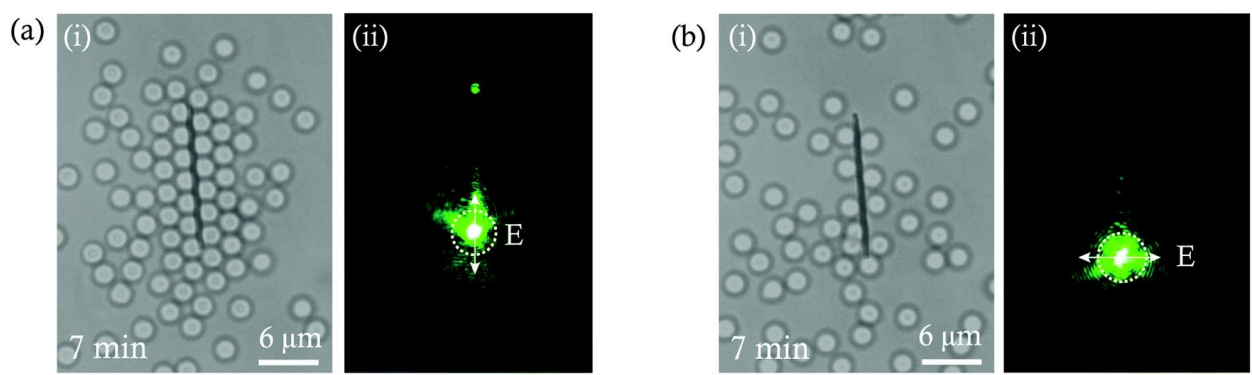

**Fig 4**: Polarization-dependence of colloidal assembly around a silver nanowire. (a) (i) Assembly of the colloids after 7 min when the polarization of the input laser is along the length of the silver nanowire. (ii) Excitation of the silver nanowire. The dotted circle indicates the laser excitation point. (b) (i) There is no significant assembly of the colloids, though few particles are trapped near the excitation point. (ii) Excitation of the wire when the polarization is perpendicular to the long axis of the wire. Reproduced with permission.[36]

This motivated us to study[36] the interaction of dielectric colloids with silver nanowires, which are illuminated at appropriate wavelengths. Such wires are prepared using chemical synthesis and are about 10 to 12 microns in length, and their thickness is about 300 to 350 nanometers. To create a delocalized heat source on the wire, a focused laser beam at 532nm was employed. This excitation created delocalized plasmons when the polarization of the excitation was aligned along the length of the nanowire. Of course, one can also change the polarization to a perpendicular orientation, which reduces the extension of the plasmon along the length of the nanowire. This specific point has interesting ramifications on the colloidal assembly as we will discuss further.

The first set of experiments that we performed was to understand the interaction of single polystyrene colloids with silver nanowires. In this experiment, the excitation polarization was aligned along the length of the wire, which created a temperature gradient. An interesting observation we made was that the colloids started to move along the length of the nanowire, and they followed the temperature gradient (from the region of low temperature to high temperature). The movement was also facilitated by the thermophoretic interaction, where the Soret coefficient happens to be negative for the polystyrene colloids.

Having explored single colloid interaction with nanowires, we then explored the case of multiple colloids (see Fig. 4). In this experiment, we first aligned the polarization of the excitation along the length of the nanowire. This resulted in the observation of the crystallization of colloids, and the symmetry was hexagonal in nature. An interesting aspect is that when we switch the polarization to the perpendicular orientation with respect to the wire, the crystallization is lost (see Fig. 4, right panel), but trapping at the location of excitation is retained. This clearly indicated that the polarization of the excitation had direct implications on colloidal dynamics and assembly. Why does this happen? As we illuminate the nanowire along its length, the temperature distribution is delocalized as the plasmons propagate throughout the length of the wire[36], whereas when the plasmon is excited perpendicular to the wire, the majority of the heating happens only at the location of excitation, and therefore, we observe the assembly happening only at that point of excitation. This explains the polarization dependence and further motivates applications where colloidal dynamics can be controlled by using the plasmonic waves in quasi-one-dimensional structures such as silver nanowires.

## Gold Microplates can extend the colloidal assembly

We have seen how plasmons in nanowires can lead to interesting dynamics and assembly of colloids. This assembly process is mainly confined to the extent of the nanowires used in the experiment, which means the scale of assembly is relatively small, say of the order of a few microns around the wire. The question is, can we extend the spatial distribution of colloidal assembly using certain plasmonic structures?

This motivated us to utilize plasmonic gold microplates,[64] as their length can be as much as a few tens of microns, whereas their thickness is of the order of a few hundred nanometers. This kind of structure can be used to create delocalized plasmon using optical techniques and hence generate optothermal potentials via temperature gradients around the excited plate. Such kind of large area structures can also be excited using evanescent optical waves. An advantage of such excitation is the minimized power budget at the sample and, hence, reduced damage.

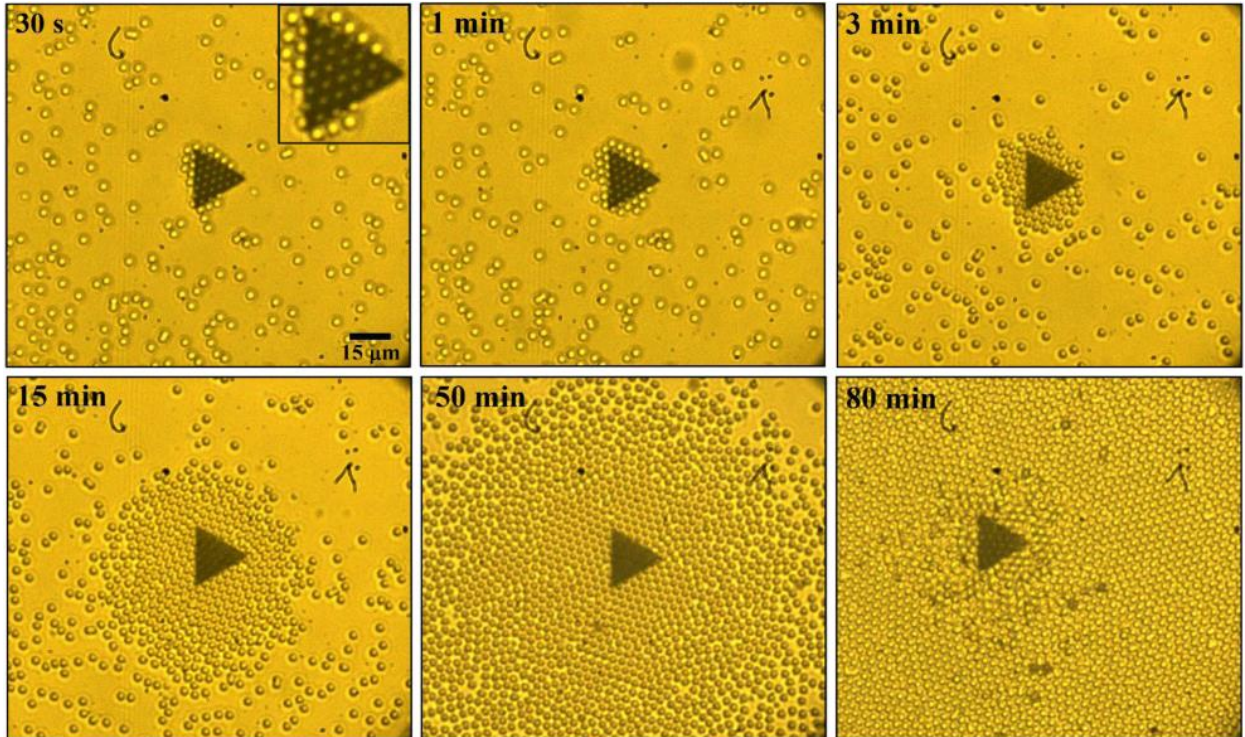

**Fig 5**: Time-series images of the gradual accumulation of the 3 $\mu$m silica particles around a gold microplate. The figure shows the snapshots of the assembly at different times, ranging from 30 s to 80 min. The inset at 30 s shows a zoomed-in image of the gold microplate. The accumulation of the particles on top of the microplate can be clearly seen. Reproduced with permission.[64]

We used gold microplates, which were chemically synthesized and had a thin layer of polymer that protected the geometry from degrading. Typical lengths were of the order of tens of microns, and the thickness was around 100 to 200 nanometers. For the assembly process, we studied silica colloids whose diameters varied from 1 to 3 microns.

For the optical excitation, we used an evanescent excitation configuration based on a dove prism. The excitation wavelength was 532 nanometers, and the maximum power used was around 300 milliwatts. Upon excitation of such microplates, we observed a very quick assembly of silica colloids within 20 to 30 seconds. A large area of hexagonal packing of colloids was observed, and their extent was as large as 300 microns (see Fig. 5). The forces at play in this interaction have contributions from optical, thermophoretic and convective forces. The optical excitation essentially comes from the evanescent waves and the thermophoretic interaction is mainly driven by the absorption of light by the microplate. This absorption further leads to temperature gradients because the plasmon is also delocalized in nature, which further facilitates a spatial extent of heat in the local vicinity of the microplate. The numerical simulations in our study[64] also indicated that the temperature increment in such structures can be as large as 7 to 8 Kelvin, which is significant for optical excitation. Interestingly, there is also a strong polarization dependence on the assembly of colloids. Most of the experiments were performed with evanescent p-polarization, where the assembly was relatively large and of the order of a few hundred microns. When we switch the polarization to s-polarized excitation, the assembly was mainly confined to the vicinity of the geometry of the plate thus clearly indicating dependence on excitation polarization. Thus, metallic microplate structures, when illuminated at an appropriate wavelength, can be harnessed for the large area optothermal assembly of colloids.

# EMERGING PROSPECTS

The interface of colloidal matter and optical fields has a rich background.[65,66] Understanding their interaction with the inclusion of thermal effects can lead to emergent prospects, which can enrich our knowledge and lead to some interesting applications. Below are a few chosen prospects that hold promise:

## Driven Soft Matter: Active Brownian Colloids

Influencing the assembly and dynamics of active Brownian colloids has turned out to be a major research area[67–71,58,72], given that active Brownian particles can emulate certain biological systems and can be harnessed as micro-scale robots[73,74]. It has opened up new avenues for the manipulation of micro-scale objects. A variety of potentials have been utilized in order to influence the assembly and dynamics of such active Brownian colloids. Generally, two categories of potentials are used based on the confinement property. For example, if one uses physical obstacles, it creates a hard potential that further influences the interaction. In contrast, one can use soft potentials such as optical tweezers[75–77] that can be used to create milder interactions with the active Brownian particles.

In recent years, there has been a variety of methods that are used to influence the dynamics and propulsion of active Brownian colloids. The challenge lies in the fact that active Brownian colloids[71] are not similar to conventional colloids because many of them have an inhomogeneous distribution of dielectric material or, in certain cases, metallic materials that are deposited on them. The external force that is acting on these particles is highly anisotropic and sensitively depends on the geometrical features and distribution of material on the colloid.

Therefore, the challenge of manipulating such active colloids needs specialized techniques. We envisage that there is scope for exploration in this area. There are already concerted efforts, such as light-controlled self-propulsion[78] and control of motility of active colloids by using ionic liquids-based phase transitions[76]. Such kind of methods clearly show that motility itself can be further used to trap objects in the vicinity of the interaction. How optical and optothermal fields participate in such interactions is a question that needs greater attention because multiple forces are at play, and the complexity increases. Such complexities can also provide new opportunities that need further exploitation.

## Harnessing Soft-Photonics

The interaction of soft matter, such as light-absorbing colloids, with optical potentials can lead to three categories of forces. The first category is the optical force, which encompasses gradient, scattering and optical spin-curl forces[11,27,79,80]. The second category is the thermophoretic force[81,82], which mainly arises due to the gradient of temperatures. The third category is the hydrodynamic force[83,84], which is mainly the result of temperature gradients coupled to the surrounding fluid in the interaction.

An interesting question to address is how parameters of light, such as intensity, phase, polarization etc, can influence all these three categories of forces in free-form optical interaction, and how they extrapolate in the presence of photonic and plasmonic devices. There is already some indirect evidence of this influence[61], in the form of large-area plasmonic traps in which there is a clear connection between the excitation polarization and the assembly of the colloids, mainly because the heating can be affected due to the polarization-dependent plasmon excitation. Such kind of polarization dependencies have also been observed[35] in structured optical traps interacting with light-absorbing colloids, where the orientation of pairs of colloids can be influenced by rotating the polarization of the trap.

In the context of the optical phase affecting dynamics, there has been significant attention in recent times[14,85]. Phase gradients have emerged as an important component in structured optical trapping. This is because phase gradients can redirect radiation pressure and facilitate transverse optical forces with respect to the optical axis of the trapping beam. Interestingly, they can also facilitate non-conservative forces[86], which can manifest in interesting ways to assemble and manipulate Brownian colloids. Another important factor related to phase gradients is that they can be coupled to polarization gradients[14,87], and therefore, the interaction between light-absorbing colloids and polarization can be enhanced or suppressed depending upon the isotropy of the medium of the colloid and the surrounding solvent.

All these 'photonic' prospects combined with soft-matter systems motivate interesting questions to be addressed, and indeed, there is scope to systematically explore the way optothermal interaction can be influenced by tailoring the optical parameters of the excitation source.

## Exploring Hot-Brownian Motion

Hot Brownian motion has emerged as an interesting concept[88], where a colloid is considered in a prototypical non-equilibrium condition. Such a system out of equilibrium can be used for a variety of effects, including trapping[89], confinement, fluid manipulation[90], assembly[59] and, as a probe of local temperature.[91,92] Another interesting application is the steering of hot swimmers[93], where the concept of photon nudging is explicitly utilized to steer swimming objects with a specific rationale that can be used to study active matter dynamics. The assembly of collective systems can also be realized using hot Brownian motion, given that it can emulate biological systems. In recent years, hot Brownian motion has also been harnessed to study quantum materials such as optically levitated nanodiamonds[94], where the local temperature can have a drastic effect on the macroscopic state of the levitated object.

In all these studies, most of the interaction is between a light-absorbing colloid or a particle and an external illumination. It would be interesting to explore the effect of structured light on hot Brownian motion and the resultant dynamics due to such interactions. Another aspect is to look into the composition, shape and size of colloids or particles that undergo hot Brownian motion and how they can be further harnessed for a variety of applications. Finally, the structure and dynamics of bio-macromolecules in temperature gradients[89] are of very high relevance in clinical characterization and analysis of disease. To this end, some emerging experiments[89] have shown great promise, and there is much scope in this direction.

# CONCLUSION

Historically, Brownian motion started as an intriguing observation[95], but it has evolved into one of the most important concepts in physics. Understanding the implications of Brownian motion due to external perturbation has turned out to be an important task, not only in statistical physics but also in allied research areas. Among the many perturbations, the optical coupling to Brownian motion can lead to many interesting aspects that can be controlled and adapted to study complex systems. The connection between Brownian motion, optical fields and the resultant optothermal effects has now found applications in a variety of scenarios, including thermoplasmonics[38] and nanophotonics[96]. It would be difficult to predict the future, but we envisage that understanding Brownian dynamics of molecular and colloidal systems can deepen our understanding of the mesoscopic world and can further lead to unanticipated applications. There is significant scope for innovation at the mesoscale that can be realized by studying Brownian colloids and their interaction with optical and optothermal fields.

## ACKNOWLEDGEMENTS

The author thanks his group members (past and present), and his colleagues - Apratim Chatterji, Vijaykumar Chikkadi and Shivprasad Patil for various discussions over the years. He also acknowledges financial support from the Swarnajayanti fellowship grant (DST/SJF/PSA-02/2017–18) from the DST-India and AOARD grant (FA2386-23-1-4054).